# *Rosetta* et après: le futur de l'exploration des comètes

Jacques Crovisier
[Observatoire de Paris]

En 1877, Jules Verne publie son roman *Hector Servadac*. Une comète frôle la Terre et en emporte un fragment avec ses habitants qui vont y vivre deux ans. C'est l'astuce trouvée par Jules Verne pour nous faire explorer avec eux une comète et le Système solaire [1]. Ce qui n'était qu'une spéculation de science-fiction, la sonde *Rosetta* l'a fait en restant près de trois ans au voisinage de la comète 67P/Churyumov-Gerasimenko (67P/C-G pour faire court) ! Coïncidence purement fortuite, il est remarquable que Palmyrin Rosette, l'astronome qui met le pied sur une comète dans la fiction de Jules Verne, et *Rosetta*, la très réelle sonde spatiale exploratrice de comète de l'Agence spatiale européenne, portent le même nom.

Dans les années 80, on pensait que les comètes étaient des boules de neige sale, et qu'elles étaient des fossiles témoins de la formation du Système solaire. C'est sur la base de ces idées que la mission *Rosetta* s'est préparée. Qu'en est-il aujourd'hui ? Où est la neige ? Qu'a-t-on appris ? Avec 67P/C-G, avait-on choisi le bon fossile ? Avait-il gardé son état de fraîcheur d'origine ? Comment poursuivre les investigations de *Rosetta* ? On peut se poser ces questions.

Mike A'Hearn (1940-2017) nous a laissé peu avant sa disparition ses impressions sur les résultats de *Rosetta* et l'avenir de la recherche cométaire [2]. Ce texte s'en est en partie librement inspiré, mais notre bilan est forcément provisoire et subjectif. Pour informations complémentaires, voir [3, 4, 5, 6].

### *Rosetta* et *Philae*

*Rosetta* a été lancée le 2 mars 2004, mais elle n'a rejoint sa comète qu'en 2014. Alors a commencé une succession d'orbites complexes dans le faible champ de gravitation de la comète. De mars 2014 à décembre 2016, les lecteurs de *L'Astronomie* ont pu suivre mois après mois les péripéties de cette mission, relatées dans la chronique de Janet Borg et d'autres articles. De cette saga, on retient quelques moments forts : le réveil de *Rosetta* le 24 janvier 2014 après une hibernation de deux ans et demi ; la surprise causée par la forme du noyau de la comète, révélée en juillet 2014 [Fig. 2 de l'article de Thérèse Encrenaz] ; les aléas de l'atterrissage du module *Philae* et de son devenir ; la désorientation de la sonde lorsqu'elle s'est trouvée prise dans des nuages de particules qui ont trompé ses senseurs stellaires le 28 mars 2015 ; l'arrêt volontaire de la mission le 30 septembre 2016.

Ainsi le 12 novembre 2014, l'atterrissage de *Philae* a été vécu en direct. Après les premiers moments passés dans l'euphorie, il fallut se rendre à l'évidence : le module *Philae*, après avoir ricoché, s'est posé de travers à un endroit imprévu. Incapable de recharger correctement ses batteries, ni de communiquer convenablement, ni de prélever par forage un échantillon du noyau, *Philae* pourra cependant remplir une partie de sa mission. Le lieu de sa chute ne sera connu qu'à la fin de la mission.

**L'évolution de la comète**

Les précédentes missions spatiales cométaires étaient toutes des missions de survol où la rencontre s'est faite à grande vitesse – environ 70 km/s pour la comète de Halley avec les sondes *VEGA* et *Giotto*, de 6 à 17 km/s pour les autres comètes. À ces vitesses, les sondes spatiales ne disposent que quelques heures d'observation utile, qui nous donnent une vision instantanée de la comète à un moment particulier de sa vie.

*Rosetta* a eu le privilège d'observer la comète 67P/C-G sans interruption de juin 2014 à fin septembre 2016 (Fig. 1). La distance au Soleil a alors diminué de 3,95 ua (lorsque la comète a été acquise après le réveil de la sonde) à 1,24 ua (au moment du passage au périhélie de la comète le 13 août 2015), puis a augmenté jusqu'à 3,83 ua le 30 septembre 2016 (fin de la mission). Son activité, que l'on peut quantifier par la production d'eau provenant de la sublimation des glaces du noyau, a varié de plus d'un facteur mille. Et des sursauts sporadiques d'activité ont été observés, non pas globalement comme on le faisait par des observations à partir de la Terre, mais localement, en identifiant les sources discrètes de la surface du noyau qui en sont responsables. *Rosetta* a pu étudier en détail les différents cycles réglant l'activité de la comète :

[Figure 1]

● **Le cycle jour/nuit :** Il est gouverné par la rotation du noyau de la comète, qui a une période de 12,4 heures. La température de la surface peut passer brusquement de -250°C la nuit à 50°C le jour, provoquant la sublimation de la glace d'eau. Cependant, des gaz gelés comme le monoxyde et le dioxyde de carbone sont beaucoup plus volatils : On peut encore en observer la sublimation du côté nuit.

● **Le cycle orbital :** La comète revient tous les 6,4 ans. Son orbite est très excentrique et sa distance au Soleil varie de 1,24 à 5,7 ua. La variation du chauffage solaire fait considérablement évoluer la production d'eau, de quelques milliers de kilogrammes par jour (à une distance de 3,9 ua du Soleil, lorsque *Rosetta* a rejoint la comète), jusqu'à environ 50 000 tonnes par jour aux alentours du périhélie (Fig. 2).

[Figure 2]

● **Le cycle des saisons :** Il est déterminé par l'inclinaison et l'orientation de l'axe de rotation. Suivant ce cycle été/hiver, différentes régions du noyau sont préférentiellement exposées au Soleil au cours de l'orbite. De la forme très bizarre du noyau de cette comète résulte une évolution complexe de l'ensoleillement et de l'activité au cours des saisons.

L'exploration de la comète de Halley et les survols qui ont suivi avaient révélé que les noyaux cométaires, loin d'être des boules de neige, même sale, étaient très noirs et ne montraient que de rares endroits où de la glace était exposée. Ce qu'a confirmé *Rosetta*, qui a en outre étudié l'évolution des plaques de neige/glace : il se produit une recondensation partielle de l'eau au cours du cycle jour/nuit, tel le givre qui se dépose sur les pare-brises de nos voitures.

● À ces cycles réguliers se superposent les **sursauts d'activité**, aussi brefs que violents. *Rosetta* en a observé à plusieurs reprises. Ils sont vraisemblablement liés à des effondrements et des fracturations de la surface du noyau.

● **L'évolution de la surface** : des galets qui roulent, des dunes qui se déplacent des collines et des falaises qui s'effondrent, des fractures et des craquelures qui se développent. Tout ceci a été observé sur une échelle de temps de deux ans (Fig. 3). Quand on songe que cette comète a effectué de nombreux passages autour du Soleil, on réalise que la surface de son noyau a perdu toute sa primitivité. Au point que l'on y recherche en vain les traces de ces cratères d'impact qui constellent la plupart des autres petits corps.

[Figure 3]

**La densité du noyau**

*Rosetta* nous a révélé la densité du noyau de 67P/C-G : 0,53 g/cm3. Cette valeur est sans surprise, c'est bien ce que l'on attendait. Mais toutes les estimations antérieures étaient indirectes et entachées d'une importante imprécision. Avec *Rosetta*, la masse du noyau de 67P/C-G a pu être évaluée avec précision à partir de l'évolution de la sonde dans le champ de gravitation de la comète. De même, sa taille et sa forme sont très bien connus. Donc, le noyau est un objet poreux. Contient-il des cavernes ou des micro-pores ? La question reste ouverte. Il se classe parmi les objets solides les moins denses du Système solaire. On note que cette densité est bien inférieure à celle des chondrites carbonées (environ 1,5 g/cm3), ces météorites que l'on présente parfois comme des résidus de noyaux cométaires.

**L'eau cométaire**

L'origine de l'eau terrestre est l'enjeu d'âpres débats. Cette eau aurait-elle été apportée par la chute de petits corps sur notre planète, en particulier de la chute de comètes ? La réponse semblerait facile à établir en tirant parti des rapports isotopiques des corps célestes, tout comme l'étude de l'ADN permet de tracer l'origine et l'évolution des êtres vivants. Il suffirait donc de comparer le rapport isotopique D/H des océans terrestres – à savoir 0,15 millième – à celui de l'eau contenue dans les comètes, astéroïdes et autres météorites.

Le problème, pour les comètes, est que ce rapport est excessivement difficile à mesurer. Il n'a pu être estimé avec une précision raisonnable que pour très peu de comètes, et les résultats ont successivement soufflé le chaud et le froid (Fig. 4). Les premières comètes mesurées – les trois « H » : Halley, Hale-Bopp et Hyakutake – avaient toutes un D/H deux fois plus élevé que le rapport terrestre. La cause semblait entendue, bien que ces trois comètes proviennent toutes du nuage de Oort alors que les comètes qui percutent la Terre seraient plutôt en majorité des comètes de la famille de Jupiter. Mais les observations du télescope spatial Herschel ont donné ensuite un rapport terrestre pour 103P/Hartley 2, une comète de la famille de Jupiter, et un rapport à peine supérieur pour C/2009 P1 (Garradd). Puis nouveau rebondissement, les mesures du spectromètre de masse de *Rosetta* ont aboutit à une valeur bien supérieure (5,3 millièmes, presque quatre fois la valeur terrestre) pour 67P/C-G, comète de la famille de Jupiter. Qu'en conclure ? Le rapport D/H cométaire est indubitablement dispersé et sa distribution est encore mal établie. Le problème reste ouvert.

[Figure 4]

**La composition de la comète**

Le bilan de la composition des glaces cométaires établi à partir de leur sublimation par le spectromètre de masse de *Rosetta* nous révèle de nombreuses nouvelles espèces chimiques observées, multipliant nos connaissances provenant de la spectroscopie à distance. Deux importants résultats sont la mise en évidence des molécules $N_2$ et $O_2$. Il reste à comprendre comment ces molécules très volatiles ont pu rester piégées dans les glaces cométaires. Jusqu'alors, la présence de $N_2$ était controversée, car la détection spectroscopique de l'ion $N_2^+$ sur laquelle elle est basée, pourtant proclamée dès le début du XX$^e$ siècle par Henri Deslandres, ne ralliait pas l'unanimité. La présence de $O_2$, une détection inattendue, pose problème, comme le montrent les différents scénarios contradictoires qui ont été imaginés pour l'expliquer.

La composition élémentaire des solides – particules de poussières et composante réfractaire du noyau – apparaît plus délicate à caractériser, car l'on a ici affaire à des molécules lourdes, voire des macromolécules. Outre les minéraux comme les silicates, se trouvent en effet des molécules organiques similaires à celles qui composent la houille, et que l'on retrouve également dans certaines météorites.

**La diversité des comètes**

Huit comètes (en comptant 21P/Giacobini-Zinner et 26P/Grigg-Skjellerup, survolées avec les sondes *ICE* et *Giotto* dépourvues de caméra opérationnelle) ont été explorées par des sondes spatiales, parmi les quelques milliers d'autres actuellement répertoriées. Nous savons maintenant que ce grand nombre d'objets présente une importante diversité d'orbite, de taille, de forme et de composition, probablement liée à des sites de formation différents et des évolutions différentes... Les coûteuses expéditions spatiales ne nous permettent d'explorer qu'un tout petit nombre de ces objets. Ces explorations doivent donc être complétées par des études menées avec des moyens plus traditionnels, portant sur l'ensemble de la population cométaire ; des études plus frustres, assurément, mais systématiques. Comme dans bien d'autres disciplines scientifiques, les progrès en astronomie ne peuvent pas provenir seulement de résultats ponctuels spectaculaires à la faveur d'une percée technologique et d'un financement occasionnel, mais procèdent aussi par l'accumulation patiente, austère et peu gratifiante à court terme d'observations qui viendront garnir des bases de données, et où les astronomes amateurs ont leur place.

Un exemple de cette diversité nous a été donné avec l'apparition, l'hiver dernier, de la comète C/2016 R2 (PanSTARRS) dont la composition s'est révélée atypique : une forte production de monoxyde de carbone, peu d'eau, absence dans le spectre visible des bandes habituellement caractéristiques de CN et $C_2$, mais présence d'émissions intenses de l'ion moléculaire $N_2^+$.

**Le futur de l'exploration spatiale des comètes**

Comme on pouvait s'y attendre, les résultats de la mission *Rosetta* ont posé plus de problèmes qu'ils n'en ont résolus. Il convient d'en envisager la suite. Voici donc ce qui serait judicieux d'entreprendre, par ordre de difficulté croissante :

● Retourner sur 67P/Churyumov-Gerasimenko quelques révolutions plus tard pour en évaluer les changements. Un simple survol, qui pourrait s'effectuer en complément d'une mission plus complexe vers d'autres objets, serait déjà une bonne approche. Un plus pourrait être l'observation de la scission du noyau. Elle est attendue et peut-être imminente, puisque des fractures sont déjà présentes !

● Réussir enfin un atterrissage sur un noyau pour effectuer la totalité de la mission prévue pour le module *Philae* et analyser *in situ* la matière de ce noyau. En prévoyant plusieurs sites (ou plusieurs atterrisseurs) pour tenir compte de la diversité des terrains. Un forage le plus profond possible (sans doute plus profond que les 30 cm qui étaient prévus pour *Philae*) est désirable pour percer la couche extérieure de matériau déjà altéré.

● Opérer un vrai retour d'échantillon, réfrigéré, qui conserve l'intégrité de la matière volatile. La mission *Startdust* ne nous avait rapporté en janvier 2006 que quelques grains de poussière prélevés en janvier 2004 dans la chevelure de 81P/Wild 2. Le prélèvement, effectué à la vitesse de 6 km/s (soit dix fois la vitesse d'une balle de fusil !), n'avait pas permis la préservation de la matière carbonée. Les missions de retour d'échantillon actuellement en cours (comme *Hayabusa-2* et *OSIRIS-Rex*) ou prévues ne visent que des astéroïdes (on note cependant que certaines des cibles sont des astéroïdes primitifs, présumés être analogues aux noyaux cométaires), et ne sont pas réfrigérées.

En décembre 2017, la NASA a pré-sélectionné dans le cadre de son programme *New Frontiers* la mission *CAESAR* (Comet Astrobiology Exploration SAmple Return), en compétition avec la mission d'exploration de Titan *Dragonfly* (Fig. 5). Cette mission, dont la cible affichée est la comète bien connue 67P/C-G, pourrait être lancée en 2024 ou 2025 pour nous en rapporter un échantillon réfrigéré de 100 g en 2038... si elle survit aux prochaines étapes de sélection et de financement.

● Une mission vers une comète réellement primitive, à longue période. Un survol à grande vitesse d'une telle comète, comme pour la comète de Halley en 1986 (avec une vitesse de 68 km/s), est réalisable avec les techniques actuelles. Mais un rendez-vous pose un véritable défi, car il est beaucoup plus exigeant en énergie. Et la découverte souvent tardive d'une nouvelle comète ne laisse qu'un court préavis pour l'organisation d'une telle mission. Mais ne craignons pas de reprendre le slogan qui a fleuri sur nos murs il y a juste cinquante ans : « soyons réalistes, demandons l'impossible ! ».

[Figure 5]

**Notes et bibliographie**

**Figures**

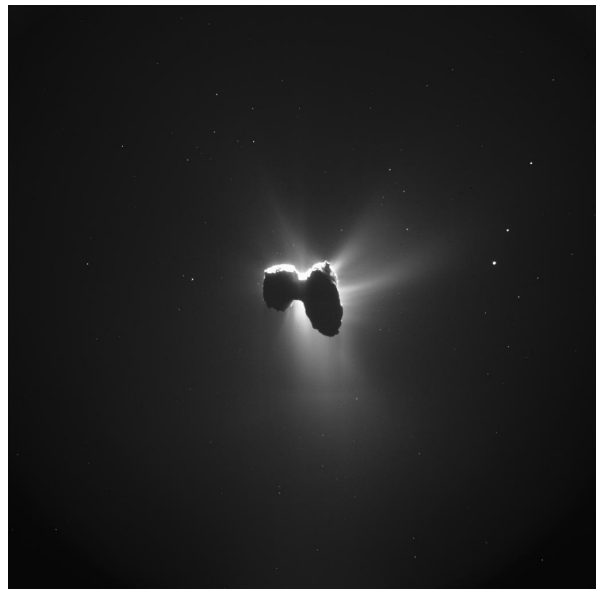

**Fig. 1** – Image de la comète 67P/C-G observée par la caméra de navigation de *Rosetta* à 329 km de distance le 27 mars 2016. Le contraste de l'image a été traité afin de mieux discerner les jets de poussières qui sont éjectés des régions actives du noyau.
(© ESA/Rosetta/NavCam)

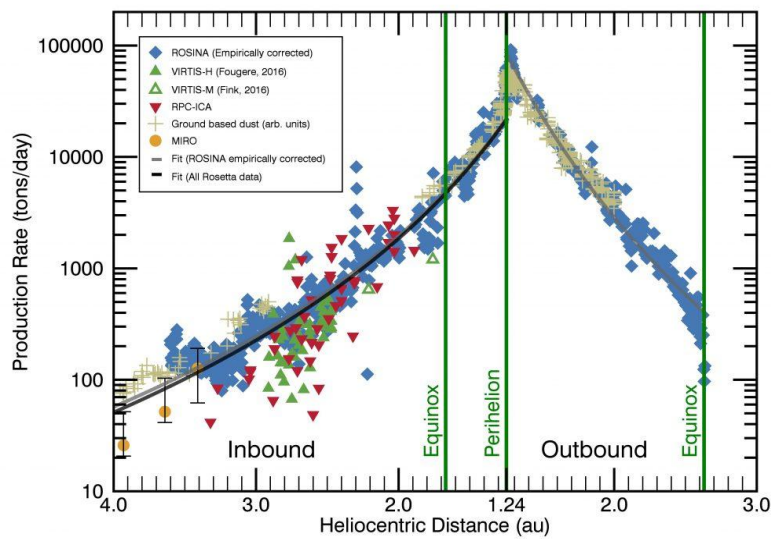

**Fig. 2** – L'évolution de la production d'eau (en tonnes par jour) de la comète 67P/C-G en fonction de la distance au Soleil, avant (à gauche) et après (à droite) le périhélie. Les résultats, obtenus par différents instruments à bord de la sonde *Rosetta* en utilisant différentes méthodes, devront encore être raccordés. (© ESA, adapté de Hansen et al. 2016, MNRAS, **462**, S491)

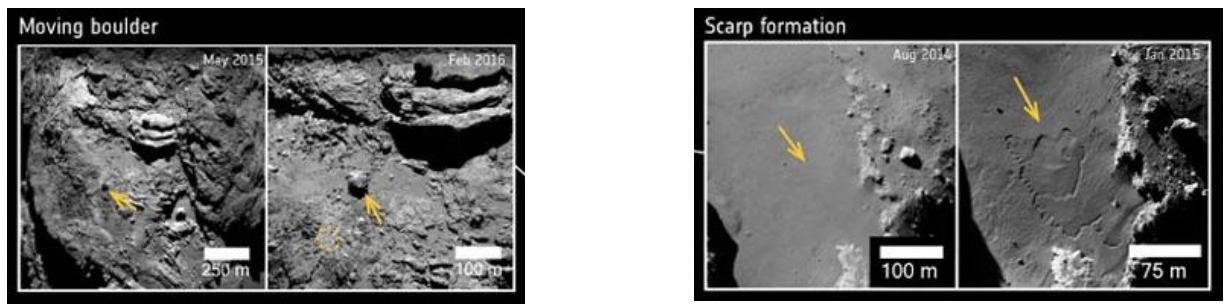

**Fig. 3** – Quelques exemples de changements observés à la surface de la comète 67P/C-G. À gauche, le déplacement sur une distance de 140 m d'un rocher d'une taille de 30 m. À droite, l'apparition de curieux escarpements sur une surface précédemment lisse.
(© ESA/Rosetta/MPS)

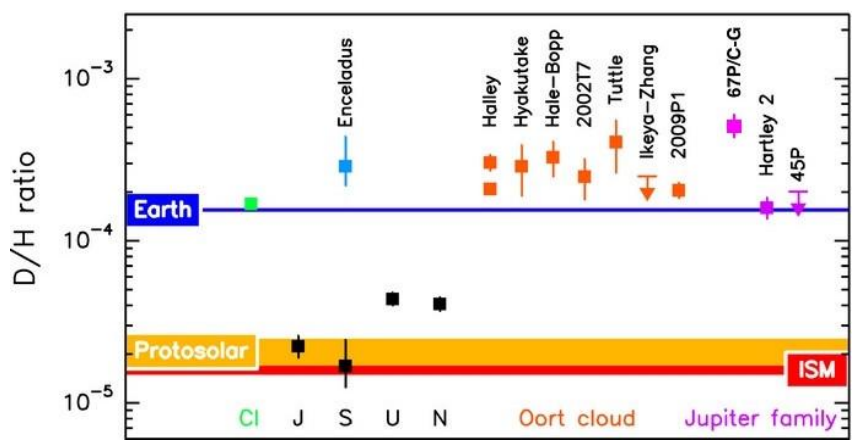

**Fig. 4** – Le rapport D/H observé dans les comètes et les autres objets du Système solaire.
(Adapté de Lis et al. 2013, *ApJ*, **774**, L3)

**Fig. 5** –
Une figure d'artiste du projet de mission de retour d'échantillon cométaire *CAESAR*. (© NASA)

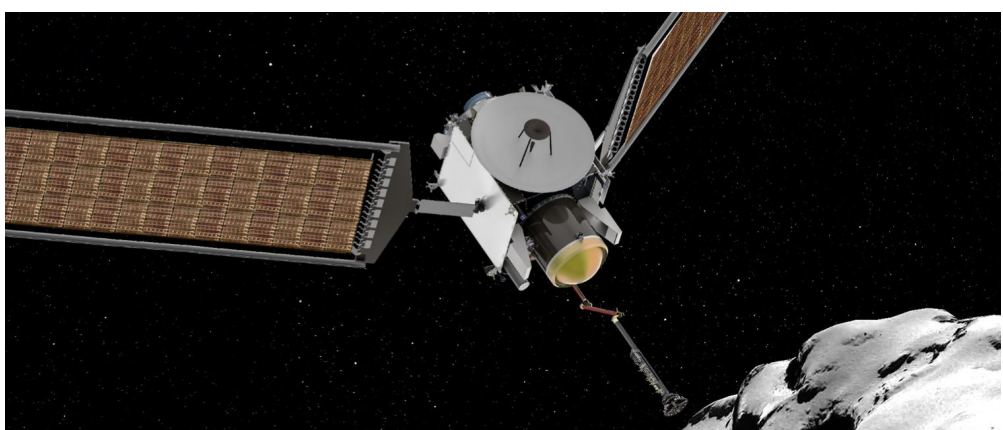